\documentclass[pdflatex,sn-mathphys-num]{sn-jnl}

\usepackage{graphicx}
\usepackage{multirow}
\usepackage{amsmath,amssymb,amsfonts}
\usepackage{amsthm}
\usepackage{mathrsfs}
\usepackage[title]{appendix}
\usepackage{xcolor}
\usepackage{textcomp}
\usepackage{manyfoot}
\usepackage{booktabs}
\usepackage{algorithm}
\usepackage{algorithmicx}
\usepackage{algpseudocode}
\usepackage{listings}
\usepackage{float}
\usepackage{lipsum}
\usepackage{geometry}
\usepackage{threeparttable}

\theoremstyle{thmstyleone}

\theoremstyle{thmstyletwo}

\theoremstyle{thmstylethree}

\begin{document}

\title[Log-Reciprocal Function in SNN Classification]{Classification Accuracy of Minimal Spiking Neural Networks Follows a Log-Reciprocal Function}

\author[1]{\fnm{Zhengdi} \sur{Zhang}}
\author[1]{\fnm{Cong} \sur{Han}}
\author*[1]{\fnm{Wenjun} \sur{Xia}}\email{wjqx@ujs.edu.cn}

\affil[1]{\orgdiv{School of Mathematical Sciences}, 
\orgname{Jiangsu University}, 
\orgaddress{\street{301 Xuefu Road}, \city{Zhenjiang}, \postcode{212013}, \state{Jiangsu Province}, \country{China}}}

\abstract{We investigate classification accuracy in minimal LIF-based spiking neural networks, examining its dependence on neuron count, stimulus nodes, and category number. Using an LLM to guide functional-form discovery, we compare power-law, exponential decay, and log-reciprocal candidates. The log-reciprocal model offers the strongest explanatory power: accuracy decays as 1/log(C), with neuron and stimulus effects marginal. This LLM-assisted approach efficiently identifies concise, interpretable descriptions, outperforming fixed-template methods. Our findings highlight AI's utility in computational neuroscience for uncovering interpretable relationships under resource constraints.}

\keywords{spiking neural networks, Leaky Integrate-and-Fire (LIF) neuron, log-reciprocal function, AI-assisted function discovery}

\maketitle

\section{Introduction}\label{sec1}

As a cornerstone of computational neuroscience and neuromorphic engineering, spiking neural networks (SNNs) offer a biologically plausible framework for understanding information processing in the brain \cite{ref3}. Among various neuron models, the Leaky Integrate-and-Fire (LIF) model strikes an effective balance between biological realism and computational tractability, making it a widely adopted building block for studying network-level dynamics and computation \cite{ref1,ref2}. A long-standing question in the field concerns the fundamental relationship between the structural parameters of a neural network and its functional capacity, particularly in resource-constrained regimes reminiscent of biological systems. Prior investigations have often focused on isolated factors: some studies have examined how classification performance varies with the number of neurons, typically observing diminishing returns \cite{ref5}, while others have analyzed the impact of input dimensionality or task complexity on network accuracy. However, a systematic exploration that concurrently varies multiple key parameters—network size (neuron count), input dimensionality (stimulus node number), and task complexity (classification category number)—in LIF-based SNNs remains limited. Understanding their combined and relative influence is crucial for revealing efficiency principles in neural computation, especially in the minimal regimes that may reflect fundamental computational units in biological circuits \cite{ref4}.

Recent advances in artificial intelligence have opened new avenues for accelerating scientific discovery. In particular, large language models (LLMs) have demonstrated emergent capabilities in reasoning, pattern recognition, and even proposing scientific hypotheses or functional relationships from complex datasets \cite{ref6}. This aligns with a growing paradigm of "human-AI collaboration" in research, where AI assists in uncovering patterns or formulating models that might be non-intuitive or labor-intensive to derive using traditional analytical methods alone \cite{ref7}. Such collaborative approaches have shown promise in fields ranging from material science to mathematical discovery \cite{ref8}.

In this work, we bridge these two strands of inquiry. We systematically investigate the classification capability of minimal LIF spiking neural networks by conducting a multi-dimensional parameter sweep over neuron count, stimulus node number, and category number. Beyond employing conventional fitting techniques (e.g., linear and polynomial regression), we leverage a large language model as a collaborative tool to aid in discovering and interpreting the underlying functional relationships that govern network performance. Our study aims to: 1) characterize the functional relationships between classification accuracy and the three core parameters, using LIF neurons \cite{ref9}, 2) identify which factor exerts the dominant influence on performance in this constrained setting, and 3) demonstrate how AI-assisted analysis can complement traditional methods to uncover concise and interpretable mathematical descriptions of complex phenomena in computational neuroscience \cite{ref10}. The findings not only elucidate fundamental constraints in spiking network performance but also contribute to the emerging methodology of human-AI collaborative scientific exploration.

\section{Methods}\label{sec2}

\subsection{Computational Model of Spiking Neuronal Networks}
We constructed spiking neural network models based on the Leaky Integrate-and-Fire (LIF) neuron model. The dynamics of each neuron is governed by the following differential equation:

\begin{equation}
\tau_m \frac{dV}{dt} = -(V - E_L) + \frac{I_{\text{noise}} + I_{\text{syn}}}{g_L} + V_{\text{stim}},
\label{eq:LIF_modified}
\end{equation}
where $\tau_m$ is the membrane time constant, $V$ is the membrane potential, $E_L$ is the resting (leakage reversal) potential, $g_L$ is the leakage conductance, $V_{\text{stim}}$ represents the external stimulus voltage, $I_{\text{noise}}$ models intrinsic physiological variability (implemented as a Poisson-distributed noise current, as detailed below), and $I_{\text{syn}}$ is the total synaptic current received from other neurons. When $V$ reaches the threshold $V_{\text{th}}$, a spike is generated, after which $V$ is reset to $V_{\text{reset}}$ and held during a refractory period $\tau_{\text{ref}}$.

\subsubsection{Noise Current Model}
The intrinsic noise current $I_{\text{noise}}$ was implemented as a Poisson-distributed stochastic process to model physiological variability. Each millisecond, an integer $k$ was sampled from a Poisson distribution with mean $\lambda = 2$, constrained to the range $0 \leq k \leq 5$ via rejection sampling. The resulting noise current, $I_{\text{noise}} = 65k$ pA, was updated every millisecond and remained constant between updates.

\subsubsection{Synaptic Current Model}
The total synaptic current $I_{\text{syn}}$ is the sum of exponentially decaying excitatory ($I_{\text{syn,E}}$) and inhibitory ($I_{\text{syn,I}}$) currents:
\begin{equation}
\frac{dI_{\text{syn,E}}}{dt} = -\frac{I_{\text{syn,E}}}{\tau_{\text{syn,E}}} + \sum_{\text{exc. spikes}} \delta(t - t_{\text{spike}}) \cdot a_{\text{ex}}
\end{equation}
\begin{equation}
\frac{dI_{\text{syn,I}}}{dt} = -\frac{I_{\text{syn,I}}}{\tau_{\text{syn,I}}} + \sum_{\text{inh. spikes}} \delta(t - t_{\text{spike}}) \cdot a_{\text{in}}
\end{equation}
where $a_{\text{ex}}$ and $a_{\text{in}}$ are the excitatory and inhibitory synaptic current increments, respectively, $\tau_{\text{syn,E}}$ and $\tau_{\text{syn,I}}$ are the excitatory and inhibitory synaptic time constants, and $\delta(t)$ is the Dirac delta function.

\subsubsection{Distance-Dependent Connection Probability and Synaptic Weight}
Structural connectivity between neurons was established probabilistically based on their Euclidean distance. The probability $p_{ij}$ of a directed connection from neuron $i$ to neuron $j$ follows a Gaussian decay, inspired by the model in \cite{ref4}:
\begin{equation}
p_{ij} = p_{\text{max}} \cdot \exp\left(-\frac{||\mathbf{x}_i - \mathbf{x}_j||^2}{2\sigma_c^2}\right)
\label{eq:conn_probability}
\end{equation}
where $p_{\text{max}}$ is the maximum connection probability, $\sigma_c$ is the spatial standard deviation (controlling the spatial extent of connectivity), and $\mathbf{x}_i, \mathbf{x}_j$ are the positions of the pre- and post-synaptic neurons, respectively.

Synaptic Weight: Upon the establishment of a connection, the strength (weight) $w_{ij}$ of the synapse from neuron $i$ to neuron $j$ is also modulated by distance, following a Gaussian decay as defined in \cite{ref4}:
\begin{equation}
w_{ij} = c \cdot \frac{1}{\sqrt{2\pi\sigma_w^2}} \cdot \exp\left(-\frac{||\mathbf{x}_i - \mathbf{x}_j||^2}{2\sigma_w^2}\right)
\label{eq:synaptic_weight}
\end{equation}
where $c$ is a scaling factor ($c = \sigma_c \times 32$, with $\sigma_c = 1000$), and $\sigma_w$ is the spatial standard deviation governing the weight distribution (typically, $\sigma_w = \sigma_c$). This ensures that physically closer neuron pairs tend to form stronger synaptic connections. Neurons were classified as excitatory (80\%) or inhibitory (20\%) and randomly distributed within the network space.

\subsection{Stimulus Input Implementation}

The stimulus input $V_{\text{stim}}$ represents external voltage pulses delivered to the network. This input is uniquely defined by a programmable binary matrix $\mathbf{S} \in \{0,1\}^{N_{\text{stim}} \times m}$, where $N_{\text{stim}}$ denotes the total number of neurons in the network that receive external stimulation (i.e., the number of stimulus nodes), and $m$ denotes the total number of defined time points within one stimulus cycle.

\textbf{Matrix Structure}:
\begin{itemize}
    \item \textbf{Row Dimension}: Each row $i$ ($1 \leq i \leq N_{\text{stim}}$) of the matrix corresponds to a specific stimulus node (i.e., a neuron receiving external input). The row vector $\mathbf{S}(i, :)$ defines the temporal sequence of pulses that this neuron will receive throughout the stimulus cycle.
    \item \textbf{Column Dimension}: Each column $k$ ($1 \leq k \leq m$) corresponds to a specific stimulus time point. The column index $k$ maps to an absolute time of $t_k = 200 \times (k-1)$ milliseconds (ms). Thus, the time points are uniformly distributed along the time axis with an interval of 200 ms, covering a total duration of $200 \times (m-1)$ ms.
\end{itemize}

\textbf{Stimulus Encoding Rule}:
A rectangular voltage pulse with an amplitude of 500 mV is delivered to neuron $i$ at the absolute time $t_k$ ms if and only if the matrix element $\mathbf{S}(i, k) = 1$. The pulse lasts for 0.5 ms. If $\mathbf{S}(i, k) = 0$, no stimulus is applied at the corresponding time point $t_k$.

\begin{equation}
V_{\text{stim}, j}(t) =
\begin{cases}
500 \text{ mV} & \text{if } \exists \, k \in \{1, 2, \dots, m\} \text{ such that } \mathbf{S}(j, k) = 1 \text{ and } t \in [t_k, t_k + 0.5] \\
0 \text{ mV} & \text{otherwise}
\end{cases}
\label{eq:stim_voltage}
\end{equation}
where $t_k = 200 \times (k - 1)$ ms.

\subsection{Stimulus Matrix Generation and Classification Design}

The stimulus matrices for each category were procedurally generated to ensure distinct yet controlled patterns for classification. For a given number of stimulus nodes ($S$), the binary matrix $\mathbf{S}_c$ for a specific category $c$ was generated as follows:

\begin{enumerate}
    \item \textbf{Position Framing:} A distinct, contiguous block of positions (rows and time columns) within the $S \times m$ matrix was allocated for category $c$. This block served as the primary encoding region for that category.
    \item \textbf{Sparsity Control:} Within this allocated block, a fixed percentage of elements (within a predefined range) were randomly set to `1`, with the rest being `0`. The specific percentage was chosen consistently across categories to maintain comparable overall input strength.
    \item \textbf{Instance Variability:} To capture pattern variation within a single category and provide robust training data, 10 distinct matrix instances were generated for each category $c$ by repeating the random `1` assignment within its defined block and sparsity constraint. These 10 instances constituted the samples for that category.
\end{enumerate}

The set of categories for an experiment was generated sequentially. The stimulus matrices for a 2-class problem ($C=2$) were generated first, defining two distinct spatial-temporal blocks. To create a 3-class problem ($C=3$), a new third category was generated by defining a new, distinct block and populating it with `1`s according to the same sparsity rule, while retaining the original two categories unchanged. This process was repeated iteratively to extend to $C=4, 5, \dots, 30$ classes, ensuring that the class structure was cumulative and directly comparable across different values of $C$.

Using this method, we generated a comprehensive stimulus set that traversed all combinations of:
\begin{itemize}
    \item Stimulus nodes: $S = 1, 2, 3$
    \item Classification categories: $C = 2, 3, 4, \dots, 30$
\end{itemize}

This stimulation protocol provides a flexible and scalable encoding framework. By adjusting the scale of the matrix $\mathbf{S}$ (i.e., $N_{\text{stim}}$ and $m$) and its internal binary patterns, we can systematically control the complexity and distinctiveness of the input signals. This allows us to investigate the influence of network size ($N$), input dimensionality ($N_{\text{stim}}$), and task complexity (number of categories $C$, corresponding to the number of unique $\mathbf{S}_c$ matrices) on the network's classification performance.

\subsection{Network Architecture and Parameters}
The fundamental network architecture maintained a physiological balance with 80\% excitatory and 20\% inhibitory neurons, consistent with cortical network organization. Network density was maintained at approximately 6000 cells/mm² across all network sizes. Each simulation comprised a 10-second stabilization period followed by 4 seconds of recording (4000 ms), with integration time steps of 0.1 ms. All results are based on 100 independent trials per parameter configuration to ensure statistical robustness.

We systematically varied three key parameters across simulation ensembles with the following constraints:
\begin{itemize}
    \item \textbf{Number of neurons ($N$):} 1, 2, 3
    \item \textbf{Number of stimulus nodes ($S$):} 1, 2, 3, with the constraint $S \leq N$
    \item \textbf{Number of classification categories ($C$):} 2, 3, 4, 5, \dots, 30 (i.e., all integers from 2 to 30)
\end{itemize}

For each parameter combination $(N, S, C)$ satisfying $S \leq N$, we designed 5 distinct simulation groups. In each group, a unique set of stimulus matrices $\{\mathbf{S}_c\}$ was generated for each category $c \in \{1, 2, \dots, C\}$ based on the given $S$ and $C$. While the stimulus matrices differed across groups, they were generated under the same parameter conditions (i.e., same $S$ and $C$), ensuring that the results capture the variability due to different input patterns and providing statistical robustness.

This full-factorial design with multiple stimulus realizations allowed us to systematically explore the independent and interactive effects of network scale, input dimensionality, and task complexity on classification performance, while accounting for the variability in input encoding.

\subsection{Classification Task and Performance Evaluation}

Following the generation of stimulus matrices and network simulation, the resulting spiking activity was processed for classification analysis. For each stimulus presentation, the network's response was captured as a binary spike matrix $\mathbf{R} \in \{0,1\}^{N \times T'}$, where $N$ is the number of neurons and $T'$ is the number of recording time steps. An element $\mathbf{R}(i, \tau) = 1$ indicates that neuron $i$ fired a spike at time step $\tau$.

To extract compact and informative features, we performed \textbf{coordinate compression} on each output spike matrix $\mathbf{R}$. This process involved identifying all matrix coordinates $(i, \tau)$ where $\mathbf{R}(i, \tau) = 1$, representing the spatial (neuron index) and temporal (time bin) locations of each spike event. These coordinates were then aggregated and binned across the population and over time to form a fixed-dimensional feature vector for each trial.

Classification was performed using \textbf{linear logistic regression} (multinomial logistic regression for multi-class problems). For a $C$-class problem, the model computes a linear score for each class $c$ based on the feature vector $\mathbf{x}$:
\[
z_c = \mathbf{w}_c^T \mathbf{x} + b_c,
\]
where $\mathbf{w}_c$ is the weight vector and $b_c$ is the bias term for class $c$. The probability that a sample belongs to class $c$ is given by the softmax function:
\[
P(y = c \mid \mathbf{x}) = \frac{e^{z_c}}{\sum_{j=1}^{C} e^{z_j}}.
\]
The model was trained by minimizing a cross-entropy loss function with L2 regularization to prevent overfitting.

For each parameter configuration $(N, S, C)$, the dataset was randomly split into training (80\% of trials) and testing (20\% of trials) sets. The linear logistic regression classifier was trained on the training set and evaluated on the held-out test set. To account for variability in the data split, this training and testing procedure was repeated 10 times with different random partitions. The classification accuracy for each run was computed as the percentage of correctly classified test samples, and the \textbf{average accuracy across these 10 runs} was taken as the final performance metric for that configuration. This metric served as our primary measure of the network's classification capability.

This evaluation protocol allowed us to systematically quantify how classification accuracy varies with the number of neurons ($N$), stimulus nodes ($S$), and categories ($C$), while controlling for stochasticity in both stimulus generation and network dynamics.

\subsection{AI-Assisted Discovery of Functional Relationships}

Prior to determining the final functional form, we explored multiple traditional and machine learning approaches to characterize the relationship between classification accuracy and the three independent variables $(C, S, N)$. This exploration process was significantly accelerated and enhanced by collaboration with a large language model (LLM), specifically leveraging its capability to propose, evaluate, and refine mathematical hypotheses.

\begin{figure}[htbp]
\centering
\includegraphics[width=0.8\textwidth]{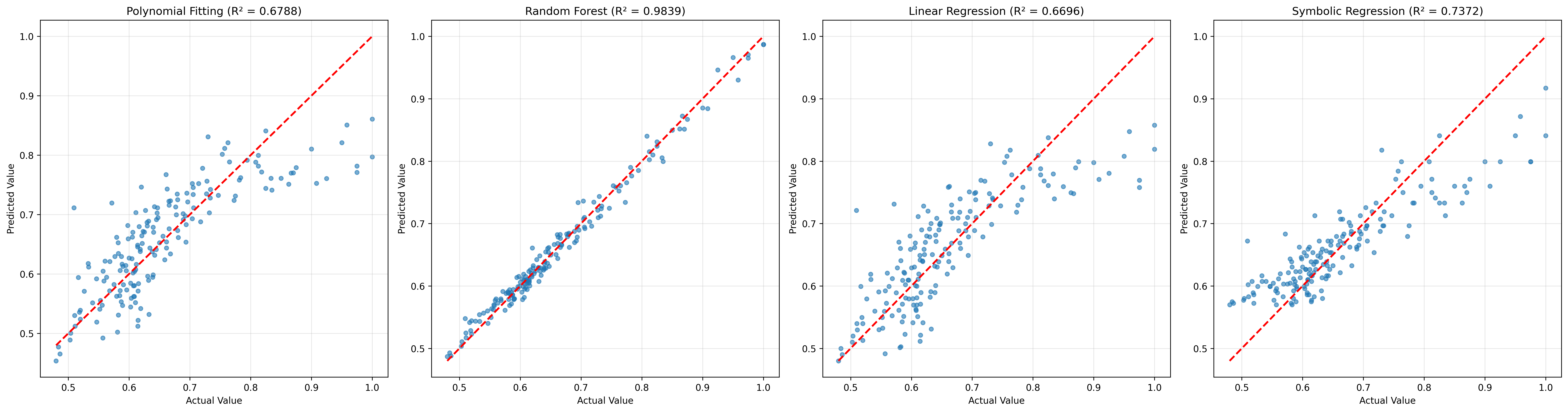}
\caption{Comparison of model predictions against actual values. The diagonal red line represents perfect prediction. All four models show varying degrees of accuracy, with Random Forest achieving near-perfect alignment.}
\label{fig:model_comparison}
\end{figure}

\begin{figure}[htbp]
\centering
\includegraphics[width=0.9\textwidth]{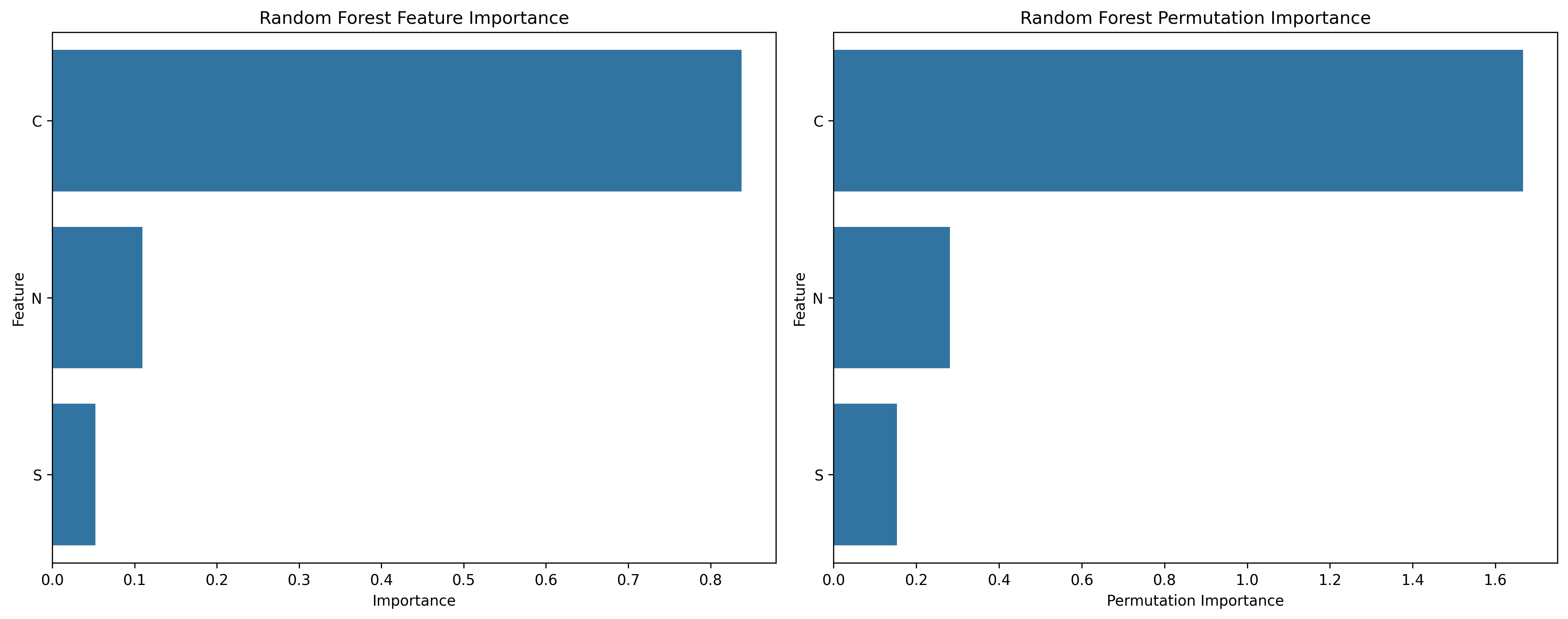}
\caption{Feature importance analysis for the Random Forest model. (Left) Standard feature importance shows $C$ (number of categories) is the dominant predictor, accounting for 83.78\% of the explained variance. (Right) Permutation importance confirms the robustness of this finding. This result is consistent with the LLM's guidance, both emphasizing the critical role of $C$ in determining the functional relationship.}
\label{fig:feature_importance}
\end{figure}

\textbf{Initial Exploration with Traditional Methods:}
We first employed several conventional regression techniques (see model comparison in Figure~\ref{fig:model_comparison}):
\begin{itemize}
\item \textbf{Linear Regression:} Simple linear models $f(C,S,N) = \beta_0 + \beta_1C + \beta_2S + \beta_3N$ yielded poor fit ($R^2 = 0.6696$), failing to capture the nonlinear relationship.
\item \textbf{Polynomial Regression:} Second-order polynomial models with interaction terms achieved moderate performance ($R^2 = 0.6788$) but lacked interpretability, with coefficients having no clear physical meaning (Equation: $f(C,S,N) = 0.9721 - 0.0107C - 0.0541S - 0.0510N + 0.0017C\cdot S - 0.0009C\cdot N + 0.0134S\cdot N$).
\item \textbf{Symbolic Regression using a Genetic Algorithm (GA):} A GA-based symbolic regression was employed to search for an optimal mathematical expression. However, this method is inherently constrained by a predefined set of mathematical symbols and operators, limiting its flexibility in discovering diverse functional forms. The discovered expression achieved moderate accuracy ($R^2 = 0.7372$) but was complex and unintuitive: $f(C,S,N) = \sqrt{\sqrt{\sqrt{(C\cdot N)^{-1}}}}$. While suggesting a potential power-law relationship between $C$ and accuracy, the specific nested radical structure offered limited mechanistic insight and was difficult to interpret meaningfully within the context of neural network dynamics.
\item \textbf{Random Forest (RF):} A non-parametric RF model achieved exceptional predictive accuracy ($R^2 = 0.9839$, RMSE = 0.0139) and robustly identified the number of categories ($C$) as the dominant feature, accounting for 83.78\% of the explained variance (feature importance analysis shown in Figure~\ref{fig:feature_importance}). However, as a black-box model, it provided no explicit functional form. Notably, the feature importance results from RF strongly aligned with the insights later refined by the LLM, both highlighting the overwhelming influence of $C$ on the functional relationship.
\end{itemize}

\textbf{LLM-Guided Hypothesis Generation and Refinement:}
The outstanding performance of the Random Forest model confirmed the dominant role of $C$ but left the exact functional relationship opaque. Similarly, while symbolic regression (GA) indicated a power-law form, its output was not interpretable. To bridge this gap, we engaged in an interactive, iterative dialogue with a state-of-the-art LLM, which effectively synthesized the strengths of both data-driven approaches: the power-law insight from symbolic regression and the clear feature dominance from Random Forest. The process involved:
\begin{enumerate}
\item \textbf{Problem Framing:} Presenting the RF feature importance results (Figure~\ref{fig:feature_importance}) and the core scientific question: "What is a parsimonious, interpretable mathematical function that describes how classification accuracy depends on task complexity ($C$), given minor corrections from network parameters ($S$, $N$)?"
\item \textbf{Hypothesis Generation:} Building on the power-law hint from symbolic regression, the LLM proposed several candidate function families based on common functional forms observed in complex systems: exponential decay ($e^{-\lambda C}$), negative logarithmic decay ($a - b \ln C$), and power-law decay ($a C^b$). These forms were subsequently fitted to the data with additive linear corrections for $S$ and $N$.
\item \textbf{Interactive Refinement:} We iteratively tested these proposals on our dataset. The LLM assisted in formulating the specific additive structure $f(C,S,N) = g(C) + h(S) + k(N)$ based on the RF insight that $C$ was dominant while $S$ and $N$ had smaller, approximately linear effects. The LLM also suggested testing both multiplicative and additive combinations of the nonlinear term with the linear terms. After evaluating the candidate forms, we examined their asymptotic behavior as $C \to \infty$ and found that none of the LLM-proposed forms satisfied the physical requirement that accuracy should vanish with unbounded task complexity. Guided by this analysis and the LLM's comparative assessments, we manually modified $g(C)$ to a log-reciprocal form, which yields a physically reasonable limit as $C \to \infty$ while preserving the overall decaying trend.

\item \textbf{Model Selection:} Quantitative comparison (Figure~\ref{fig:model_comparison}) showed that while the Random Forest achieved the highest predictive accuracy ($R^2 = 0.9839$), it lacked interpretability. Among interpretable models, the additive log-reciprocal model (Equation~\ref{eq:logreciprocal_model}) offered the best balance between performance and interpretability, providing a clear functional description of the observed dependence while maintaining reasonable predictive accuracy. This final model can be viewed as a synthesis: it incorporates the log-reciprocal dependence on $C$ (as suggested by comparative analysis of candidate forms) within a simple, additive framework that respects the relative influence of variables as identified by Random Forest.
\end{enumerate}

This collaborative, human-AI workflow allowed us to efficiently navigate a vast space of potential models. While data-driven methods like Random Forest achieved superior predictive performance and symbolic regression hinted at key functional forms, the LLM-guided analytical approach integrated these insights to lead us to a physically interpretable relationship that provides mechanistic understanding of the underlying phenomena—an insight that might have been obscured by purely data-driven black-box models.

\subsection{Model Parameters}
The complete set of model parameters is summarized in the following tables.

\begin{table}[htbp]
\centering
\caption{Neuron Model Parameters}
\label{tab:neuron_params}
\begin{tabular}{lcc}
\toprule
\textbf{Parameter} & \textbf{Symbol} & \textbf{Value} \\
\midrule
Membrane time constant & $\tau_m$ & 10 ms \\
Resting potential & $V_{\text{rest}}$ & -70 mV \\
Threshold potential & $V_{\text{th}}$ & -50 mV \\
Reset potential & $V_{\text{reset}}$ & -70 mV \\
Refractory period & $\tau_{\text{ref}}$ & 3 ms \\
Leakage conductance & $g_L$ & 10 nS \\
\bottomrule
\end{tabular}
\end{table}

\begin{table}[htbp]
\centering
\caption{Synaptic and Input Parameters}
\label{tab:synapse_params}
\begin{tabular}{lcc}
\toprule
\textbf{Parameter} & \textbf{Symbol} & \textbf{Value} \\
\midrule
Excitatory synaptic weight & $\alpha_{\text{EX}}$ & 10 \\
Inhibitory synaptic weight & $\alpha_{\text{IN}}$ & 20 \\
Synaptic time constant & $\tau_{\text{syn}}$ & 10 ms \\
Stimulus voltage amplitude & $V_{\text{stim}}$ & 500 mV \\
Stimulus pulse duration & $T_{\text{pulse}}$ & 0.5 ms \\
Inter-stimulus interval & $T_{\text{ISI}}$ & 200 ms \\
Excitatory synaptic time constant & $\tau_{\text{syn,E}}$ & 2 ms \\
Inhibitory synaptic time constant & $\tau_{\text{syn,I}}$ & 5 ms \\
Gaussian connection $\sigma$ & $\sigma_{\text{conn}}$ & 1000 $\mu$m \\
Max connection probability & $p_{\text{MAX}}$ & 0.25 \\
\bottomrule
\end{tabular}
\end{table}

\section{Results}\label{sec4}

\subsection{LLM-Guided Fitting and Comparison of Candidate Function Families}

Based on the three common functional forms proposed by the large language model—power function (PowerLaw), negative logarithmic (NegLog), and exponential decay (Exp)—we performed nonlinear regression to relate classification accuracy to the three independent variables: number of categories ($C$), number of stimulus nodes ($S$), and number of neurons ($N$). For a unified comparison, all three models adopted an additive structure, where the effect of $C$ is represented by different nonlinear functions $g(C)$, and the contributions of $S$ and $N$ are linearly superimposed:

\begin{equation}
f(C,S,N) = g(C) + p_3 S + p_4 N + p_5,
\label{eq:additive_template}
\end{equation}

where $g(C)$ takes the following forms:

\begin{itemize}
    \item Power function: $g(C) = p_1 C^{\,p_2}$,
    \item Negative logarithmic: $g(C) = -p_1 \ln(C + p_2)$,
    \item Exponential decay: $g(C) = p_1 \exp(p_2 C)$.
\end{itemize}

The fitting results are summarized in Table~\ref{tab:llm_candidate_models}.

\begin{table}[htbp]
\centering
\caption{Fitting performance of the three candidate models}
\label{tab:llm_candidate_models}
\begin{tabular}{lccccc}
\toprule
\textbf{Model} & \textbf{Formula} & $R^2$ & MAE & RMSE \\
\midrule
PowerLaw & $3.150\,C^{-0.054} + 0.0307S - 0.0464N - 2.000$ & 0.7610 & 0.0489 & 0.0624 \\
NegLog & $-0.1478\ln(C-0.1601) + 0.0307S - 0.0463N + 1.121$ & 0.7611 & 0.0489 & 0.0624 \\
Exp & $0.4848\,e^{-0.0997C} + 0.0307S - 0.0463N + 0.6067$ & 0.7577 & 0.0493 & 0.0628 \\
\bottomrule
\end{tabular}
\end{table}

The data show that the goodness-of-fit of the three models are remarkably close: $R^2$ values all lie between 0.758 and 0.761, and the mean absolute error (MAE) and root mean square error (RMSE) are also nearly identical. This indicates that, within the current experimental parameter range, all three functional forms can reasonably capture the overall trend of accuracy decreasing with increasing $C$.

However, when we shift our perspective from interpolation to extrapolation, all three models expose serious physical inconsistencies. In classification tasks, as $C \to \infty$, the accuracy $f(C,S,N)$ should asymptotically tend to $0$ (or at least converge to a negligibly small value). Examining the explicit formulas listed in Table~\ref{tab:llm_candidate_models}, we find that as $C \to \infty$:

\begin{itemize}
    \item \textbf{Power-law model:} $g(C) \to 0$, but the constant term $p_5 = -2.000$ drives the accuracy limit to a negative value, which completely violates the definition domain $[0,1]$ of accuracy;
    \item \textbf{Negative logarithmic model:} $g(C) \to -\infty$, causing the accuracy to diverge to negative infinity, which is mathematically unacceptable;
    \item \textbf{Exponential model:} $g(C) \to 0$, but the limit $p_5 \approx 0.607$ is far above $0$, implying that the network would maintain a non-zero accuracy even with infinitely many categories—this is inconsistent with any rational classifier behavior.
\end{itemize}

\subsection{Additive Log-Reciprocal Model}

As discussed above, the three LLM-proposed candidate models all suffer from unsatisfactory asymptotic behavior: as $C \to \infty$, their accuracy limits are either negative, divergent, or far above zero, conflicting with the physical expectation that accuracy should vanish when the number of categories grows without bound. To address this, we propose a log-reciprocal model, which yields an accuracy limit close to zero as $C \to \infty$ while preserving the overall decaying trend. The model adopts the same additive structure for $S$ and $N$:

\begin{equation}
f(C,S,N) = \frac{p_1}{\ln(C + p_2)} + p_3 S + p_4 N + p_5,
\label{eq:logreciprocal_model}
\end{equation}

where $C$, $S$, and $N$ denote the number of categories, stimulus nodes, and neurons, respectively, and $p_1$ through $p_5$ are fitted parameters.

\begin{figure}[htbp]
\centering
\includegraphics[width=0.8\textwidth]{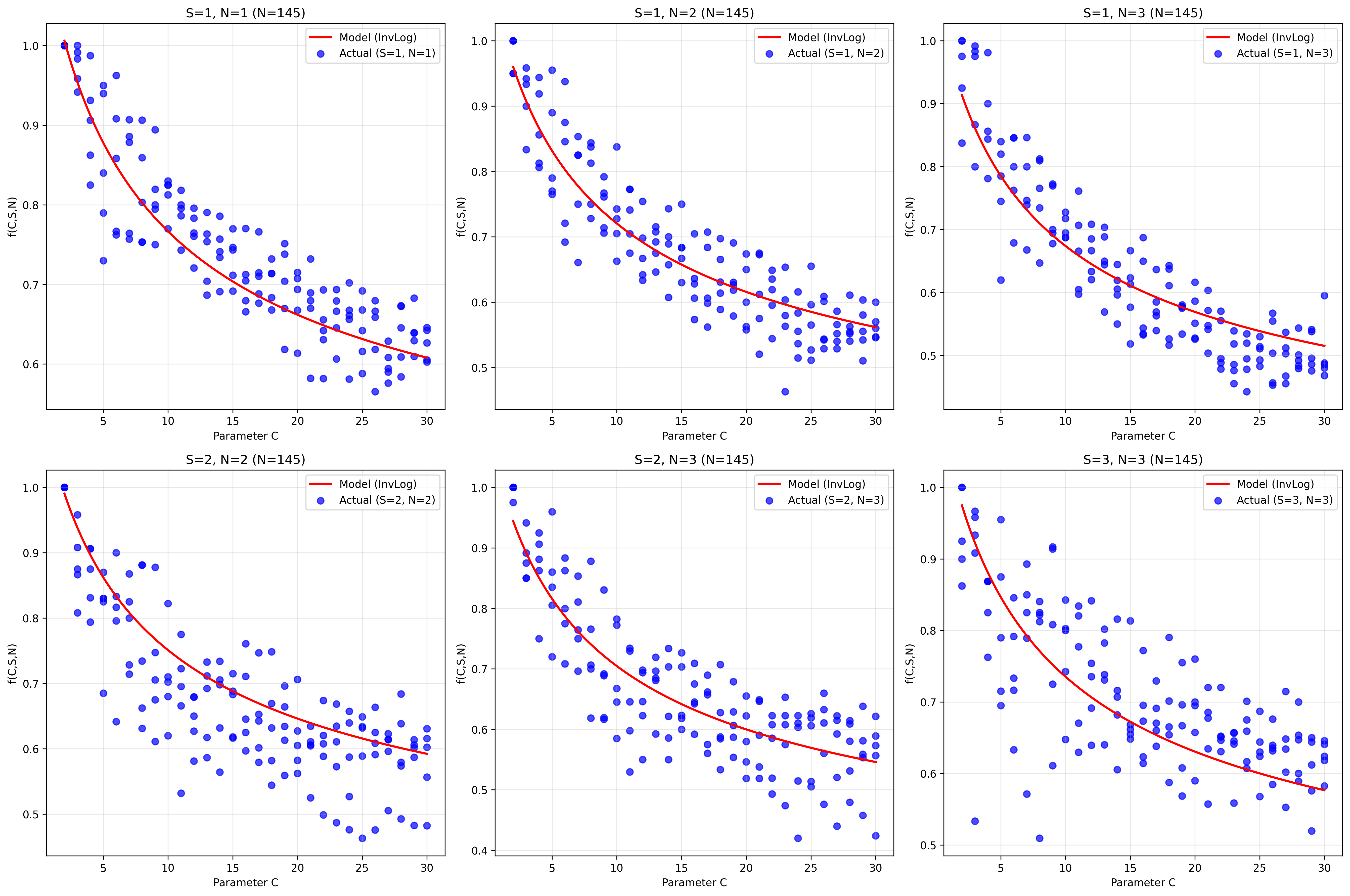}
\caption{Log-reciprocal fitting of classification accuracy as a function of category count ($C$) for various combinations of stimulus nodes ($S$) and neurons ($N$). Each subplot corresponds to a fixed $(S, N)$ configuration, illustrating the consistent log-reciprocal dependence across all network settings.}
\label{fig:sn_combinations}
\end{figure}

Nonlinear regression on the combined dataset (870 samples from five independent runs) yielded:

\begin{itemize}
    \item $p_1 = 1.910596$, $p_2 = 5.760591$, $p_3 = 0.030684$, $p_4 = -0.046346$, $p_5 = 0.089454$,
    \item $R^2 = 0.7606$, MAE = 0.0489, RMSE = 0.0624.
\end{itemize}

The fitting performance is nearly identical to that of the power-law, negative-logarithmic, and exponential models. Crucially, as $C \to \infty$, the term $p_1/\ln(C+p_2) \to 0$, leaving the accuracy limit determined solely by $p_5 \approx 0.0895$, which is close to zero. This aligns with the physical expectation that accuracy should tend to zero as task complexity becomes arbitrarily large—a property not shared by the LLM-proposed candidates. The log-reciprocal model thus offers comparable in-sample fit with superior extrapolation plausibility.

\begin{figure}[htbp]
\centering
\includegraphics[width=0.8\textwidth]{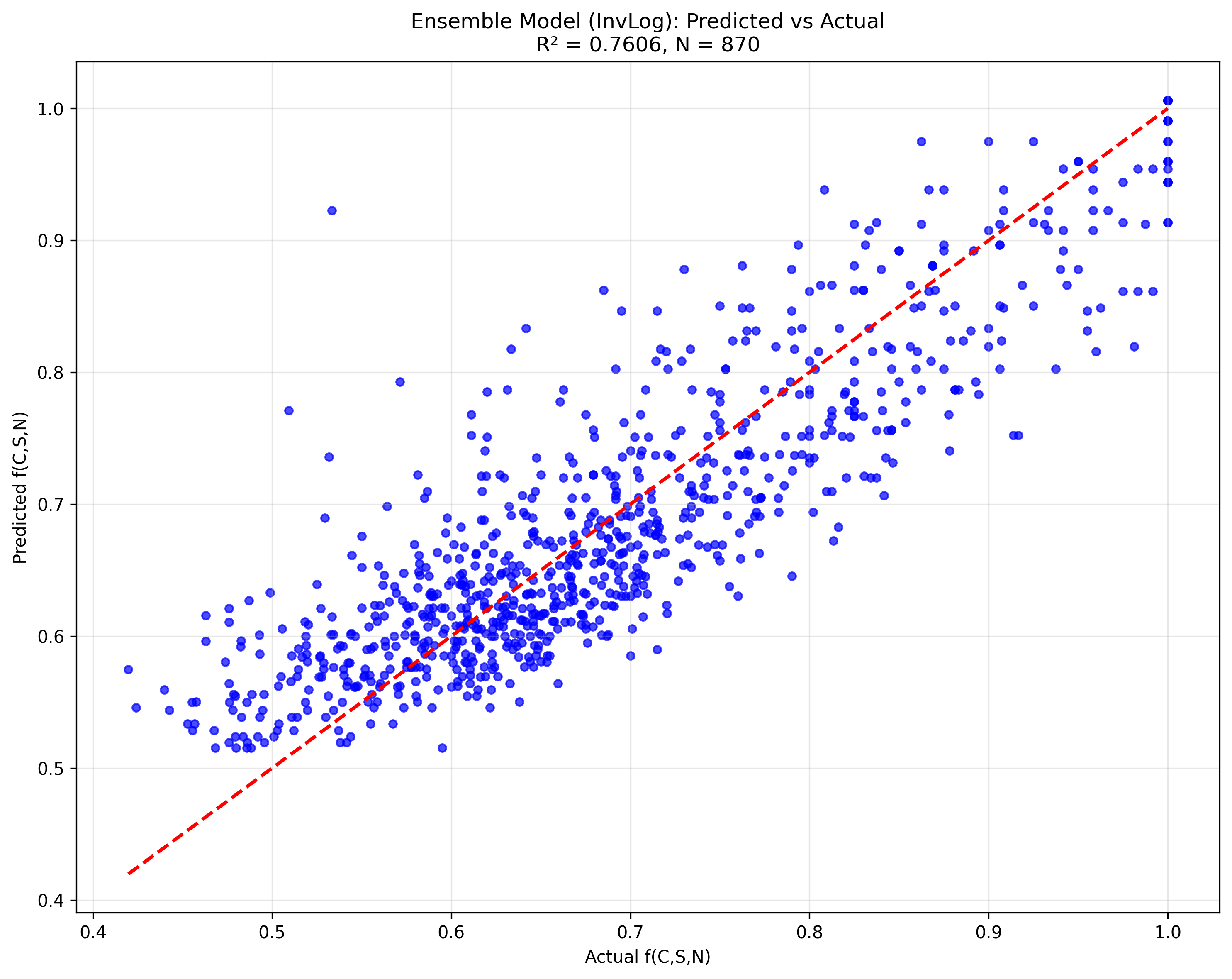}
\caption{Scatter plot of predicted versus actual classification accuracy for the log-reciprocal model. Close alignment along the diagonal demonstrates good predictive accuracy across the full value range.}
\label{fig:pred_vs_actual}
\end{figure}

Additional diagnostic analyses, including residual analysis and distribution comparisons, are provided in the Appendix.

\subsection{Control Experiment: Analysis of Stimulus Input Classification Without Neuronal Networks}

We further analyzed the log-reciprocal relationships of the form $f(C) = \ln(\alpha) / \ln(C + \alpha)$ for stimulus input classification data (direct classification on stimulus inputs without the involvement of neuronal networks), under fixed stimulus node counts $S$:

\begin{table}[htbp]
\centering
\caption{Log-reciprocal fits of classification accuracy vs. $C$ for fixed stimulus node counts (no neuronal networks involved)}
\label{tab:conditional_logreciprocal}
\begin{tabular}{lcccc}
\toprule
\textbf{Stimulus Nodes ($S$)} & \textbf{Data Points} & \textbf{Parameter $\alpha$} & \textbf{$R^2$} & \textbf{Fitting Quality} \\
\midrule
1 & 145 & 43.3207 & 0.1455 & Poor \\
2 & 145 & 95.5295 & 0.2527 & Poor \\
3 & 145 & 400.4748 & 0.2265 & Poor \\
\bottomrule
\end{tabular}
\end{table}

Table~\ref{tab:conditional_logreciprocal} shows that the log-reciprocal model yields extremely poor fits ($R^2 < 0.3$) across all stimulus node configurations. Notably, these poor fits directly indicate that there is \textbf{no valid log-reciprocal distribution between stimulus inputs and the number of categories} in the absence of neuronal networks. Fitting quality classification: Excellent ($R^2 > 0.9$), Good ($>0.7$), Fair ($>0.5$), Poor ($\leq 0.5$).

The poor fitting performance underscores two key points: first, considering only $C$ is insufficient to characterize the relationship; second, more importantly, the core conclusion that the log-reciprocal relationship does not hold for direct stimulus input classification without neuronal network participation. The log-reciprocal structure observed in network-mediated classification (Equation~\ref{eq:logreciprocal_model}) is thus not an inherent property of the stimuli themselves, but rather emerges from spiking network processing.

\begin{figure}[htbp]
\centering
\includegraphics[width=0.8\textwidth]{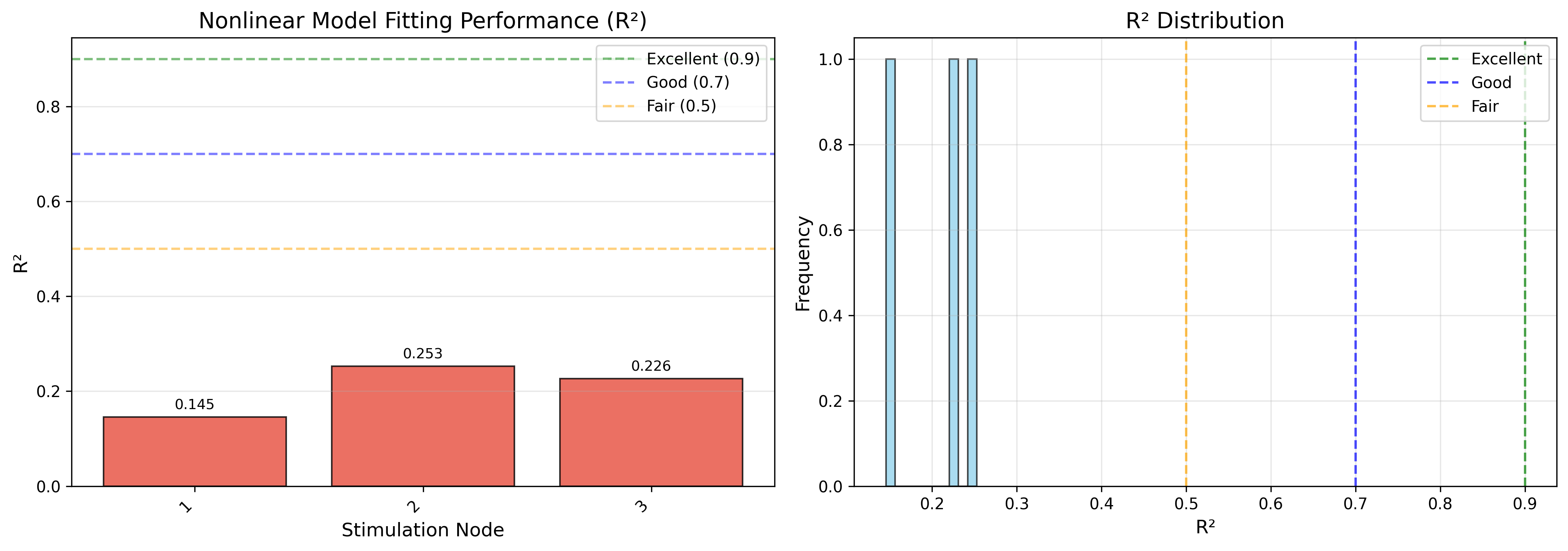}
\caption{Log-reciprocal fitting diagnostic for stimulus input classification data (no neuronal networks involved). Left: Bar chart of $R^2$ values for each stimulus node, color-coded by fitting quality thresholds (green: $R^2 > 0.9$, blue: $0.7 < R^2 \leq 0.9$, orange: $0.5 < R^2 \leq 0.7$, red: $R^2 \leq 0.5$). Horizontal dashed lines mark the 0.9, 0.7, and 0.5 thresholds. Right: Histogram of $R^2$ value distribution across all nodes, with the same threshold lines overlaid. The prevalence of red bars on the left, along with the concentration of the histogram in the low $R^2$ region on the right, confirms the overall failure of the log-reciprocal model to describe direct stimulus classification data.}
\label{fig:nonlinear_fitting_performance}
\end{figure}

\begin{figure}[htbp]
\centering
\includegraphics[width=0.8\textwidth]{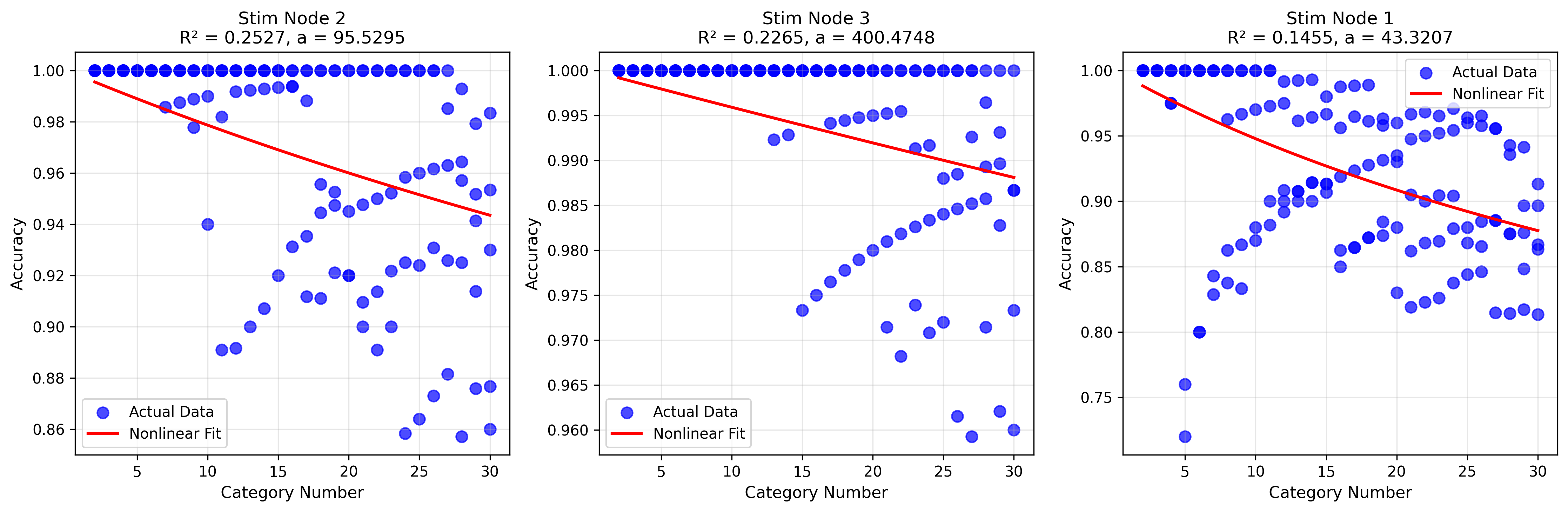}
\caption{Representative log-reciprocal fits for specific fixed parameter combinations $(S, N)$ without neuronal network involvement, with no observable log-reciprocal regularity.}
\label{fig:representative_fits}
\end{figure}

\begin{figure}[htbp]
\centering
\includegraphics[width=0.8\textwidth]{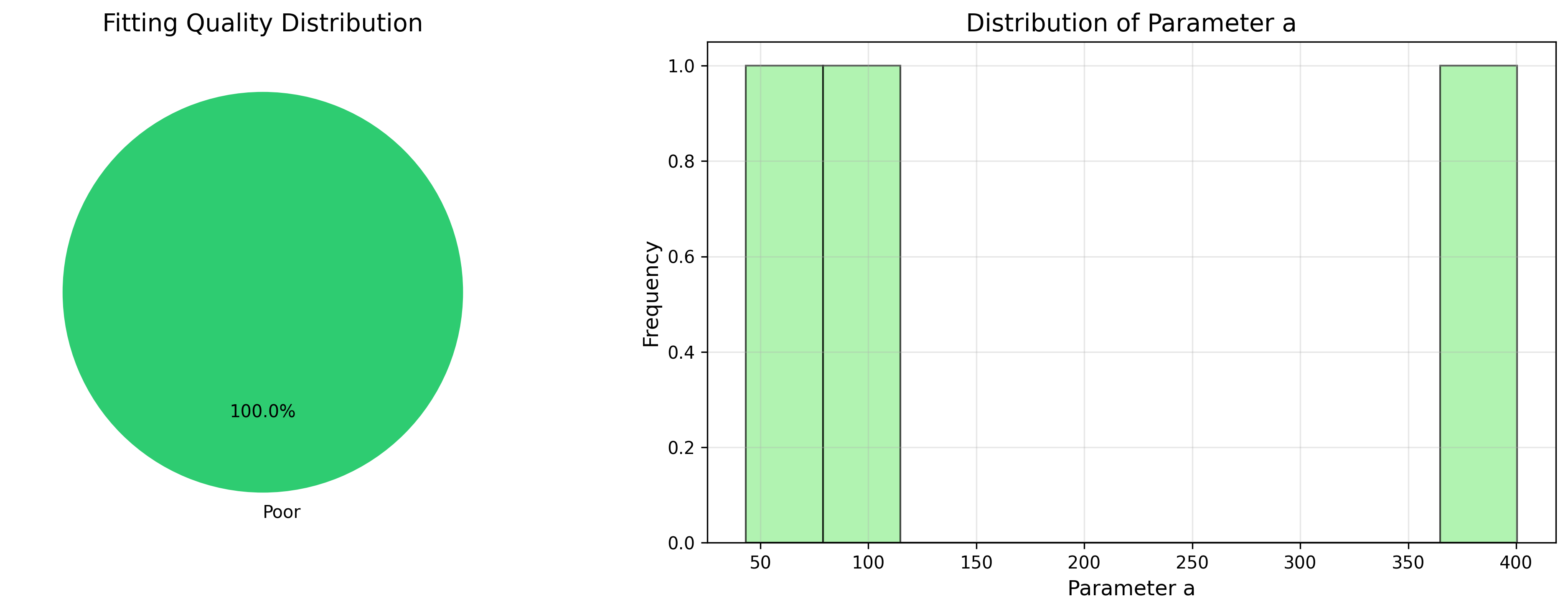}
\caption{Diagnostic summary of log-reciprocal fitting across all stimulus input parameter combinations (no neuronal networks involved). Left: Pie chart of fitting quality distribution based on $R^2$ thresholds (Excellent: $R^2 > 0.9$, Good: $0.7 < R^2 \leq 0.9$, Fair: $0.5 < R^2 \leq 0.7$, Poor: $R^2 \leq 0.5$). The predominance of the Poor sector confirms that the model $f(C) = \ln(\alpha)/\ln(C+\alpha)$ is entirely inapplicable to the stimulus data. Right: Histogram of fitted parameter $\alpha$ values across all nodes. The widely dispersed distribution indicates severe parameter instability and the absence of a unified underlying mechanism, further confirming that the log-reciprocal form fails to provide any consistent description of direct stimulus classification.}
\label{fig:fit_quality_dist}
\end{figure}

\subsection{Interpretation and Key Findings}

The additive log-reciprocal model and comparative analysis reveal several important insights:

\begin{enumerate}
    \item \textbf{Network-induced log-reciprocal relationship:} The log-reciprocal relationship between classification accuracy and task complexity is not inherent in the stimuli but emerges from spiking network processing. The control experiment (Table~\ref{tab:conditional_logreciprocal} and Figure~\ref{fig:nonlinear_fitting_performance}) shows that direct stimulus classification yields extremely poor log-reciprocal fits with $R^2 < 0.3$ across all stimulus node configurations, with all nodes exhibiting red-level fitting quality ($R^2 \leq 0.5$). In stark contrast, network-mediated classification exhibits a clear log-reciprocal relationship with $R^2 = 0.7606$. This disparity confirms that the spiking network's nonlinear transformation is the essential source of the observed functional behavior.

    \item \textbf{Dominant log-reciprocal relationship with categories:} The primary determinant of network classification accuracy is the number of categories $C$, following a log-reciprocal decay of the form $p_1/\ln(C+p_2)$. The constant term $p_5 \approx 0.0895$ is close to zero, confirming that accuracy asymptotically diminishes to zero as $C$ grows indefinitely. This aligns with the physical expectation that classification becomes increasingly difficult with more categories, and the network's information capacity is eventually exhausted.

    \item \textbf{Minor linear network parameter effects:} Stimulus node count $S$ and neuron count $N$ show small linear contributions ($p_3 = 0.030684$, $p_4 = -0.046346$). The positive $S$ coefficient suggests modest benefits from additional input channels, providing more information for discrimination; the negative $N$ coefficient may reflect overfitting or representational redundancy in extremely small networks (1-3 neurons), where additional neurons do not necessarily enhance performance.

    \item \textbf{Predominance of task complexity:} Relative coefficient magnitudes confirm that task complexity (categories) has a substantially larger impact on classification performance than network size or input dimensionality in our experimental regime. The log-reciprocal term dominates the additive model, while the linear corrections for $S$ and $N$ are marginal. This suggests that for minimal spiking networks, the fundamental limitation on performance is the number of classes to be discriminated, rather than the network's input or size parameters.
\end{enumerate}

\begin{figure}[htbp]
\centering
\includegraphics[width=0.8\textwidth]{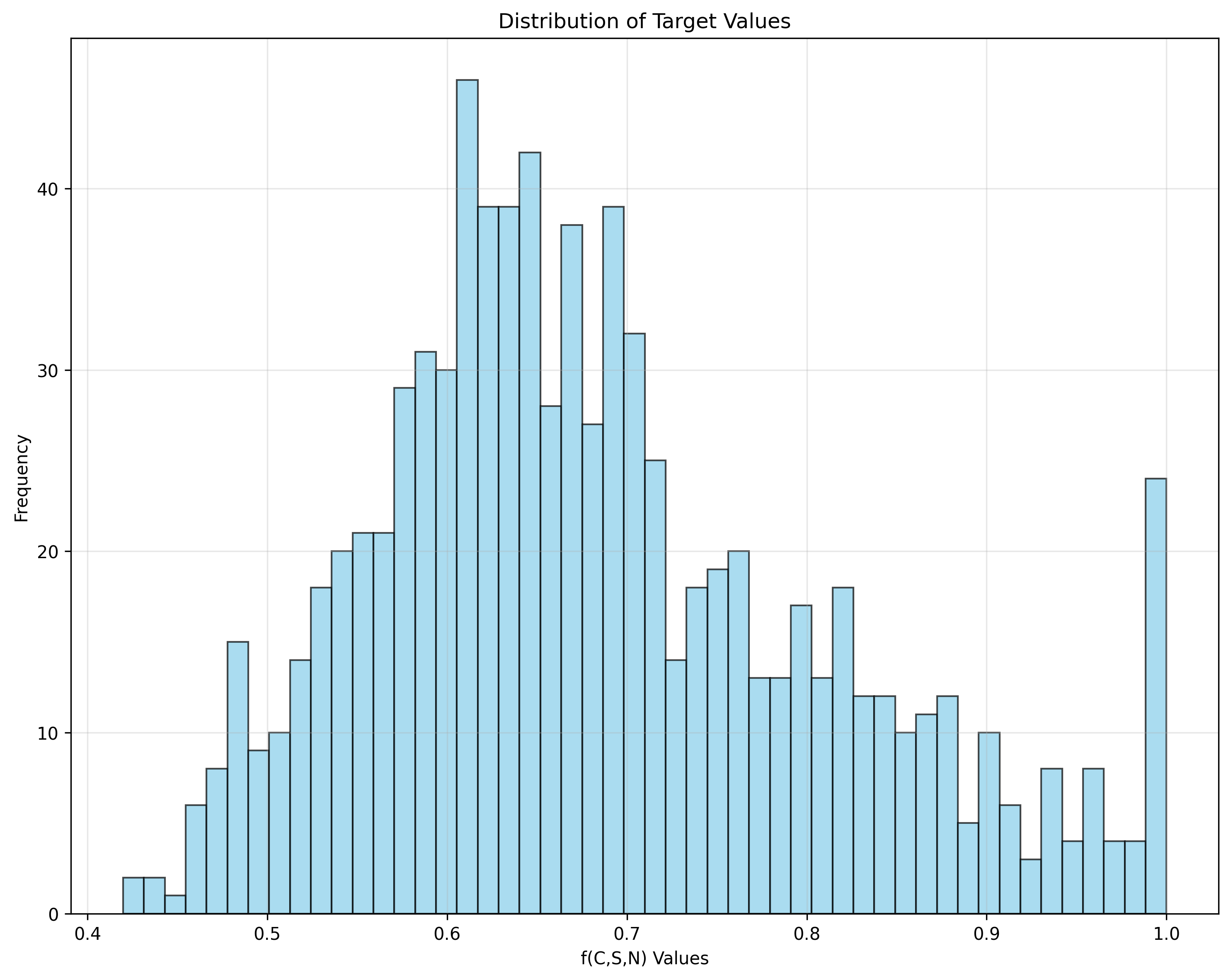}
\caption{Frequency distribution histogram of classification accuracy (represented by $f(C,S,N)$ values). The horizontal axis denotes the actual classification accuracy (i.e., $f(C,S,N)$ values), while the vertical axis indicates the frequency of occurrence for each accuracy value. The multimodal distribution feature presented in the figure intuitively reflects the variability in task difficulty across different $(S, N, C)$ parameter combinations: each peak corresponds to a concentrated region of accuracy under specific parameter configurations, meaning some parameter combinations form task groups with similar difficulty, and their classification accuracies are relatively clustered.}
\label{fig:target_dist}
\end{figure}

These results demonstrate that even in extremely small neural networks (1-3 neurons), classification accuracy follows predictable functional dependencies induced by network processing, with task complexity being the dominant factor. The additive log-reciprocal model provides a quantitative framework for understanding the interplay between network architecture and classification capability in resource-constrained regimes, and offers a physically plausible extrapolation to large-$C$ scenarios where other candidate models fail.

\section{Conclusion}\label{sec13}

This study has systematically investigated the classification capabilities of minimal spiking neural networks. Our central finding is that the classification accuracy of networks composed of only 1 to 3 neurons exhibits a precise and predictable functional dependence, which is predominantly characterized by a log-reciprocal decay with the number of task categories. The empirically derived additive log-reciprocal model, $f(C,S,N) = p_1/\ln(C+p_2) + p_3 S + p_4 N + p_5$, successfully captures this relationship, with $p_5 \approx 0.0895$ near zero, confirming that accuracy asymptotically vanishes as task complexity grows without bound—a property not exhibited by the LLM-proposed candidates.

Crucially, a dedicated control experiment demonstrated that this log-reciprocal relationship is not an inherent property of the stimulus patterns themselves—which remained highly separable under direct classification—but is a fundamental computational bottleneck induced by the spiking neural network. Direct stimulus classification yielded extremely poor log-reciprocal fits ($R^2 < 0.3$ across all configurations), while network-mediated classification exhibited a clear log-reciprocal dependence ($R^2 = 0.7606$). This underscores that the transformation of information within a dynamic, resource-constrained spiking framework imposes intrinsic limits on performance, revealing a core trade-off between task complexity and computational capacity.

Methodologically, this work highlights the value of human-AI collaboration in scientific discovery. The path to this interpretable functional relationship was significantly accelerated by leveraging a large language model as a hypothesis-generation partner. After traditional methods (linear/polynomial regression, genetic algorithms) failed to yield both accuracy and interpretability, and while a random forest model confirmed the dominant role of category number ($C$), the AI aided in efficiently exploring and refining candidate function families. Building upon the LLM's proposals—power function, negative logarithmic, and exponential decay—we manually designed the log-reciprocal form, which offers comparable in-sample fit while providing physically plausible extrapolation behavior, a critical advantage for understanding the fundamental limits of neural computation.

\subsection*{Future Perspectives and Implications}

Several follow-up directions emerge from this work, with potential but bounded relevance.

Extending Network Complexity: The current findings are restricted to 1–3 neuron networks. Testing whether the log-reciprocal form persists with balanced inhibition, structured connectivity, or plasticity would help define its domain of validity.

Theoretical Underpinnings: The empirical relationship currently lacks a mechanistic derivation. Future efforts could examine whether it arises from information-theoretic bounds or the dynamics of low-dimensional competitive systems, though such links remain to be established.

Dissecting Minor Influences: The small but significant effects of NN and SS merit further scrutiny. Controlled variations could determine whether they reflect genuine computational contributions or are artifacts of the specific parameter ranges used.

Implications for Neuromorphic Engineering: This relationship offers a quantitative benchmark for performance limits in severely resource-constrained spiking systems. It may inform early-stage design considerations, but translating it into hardware-specific guidelines will require additional calibration and validation across diverse task conditions.

In summary, this study provides a clear empirical baseline and points to concrete next steps, while its broader applicability and theoretical interpretation remain open questions.

\backmatter

\begin{appendices}

\section{Supplementary Model Diagnostics}\label{secA1}

The following supplementary figures provide additional diagnostic information about the additive log-reciprocal model's performance and residual characteristics.

\begin{figure}[H]
\centering
\includegraphics[width=0.8\textwidth]{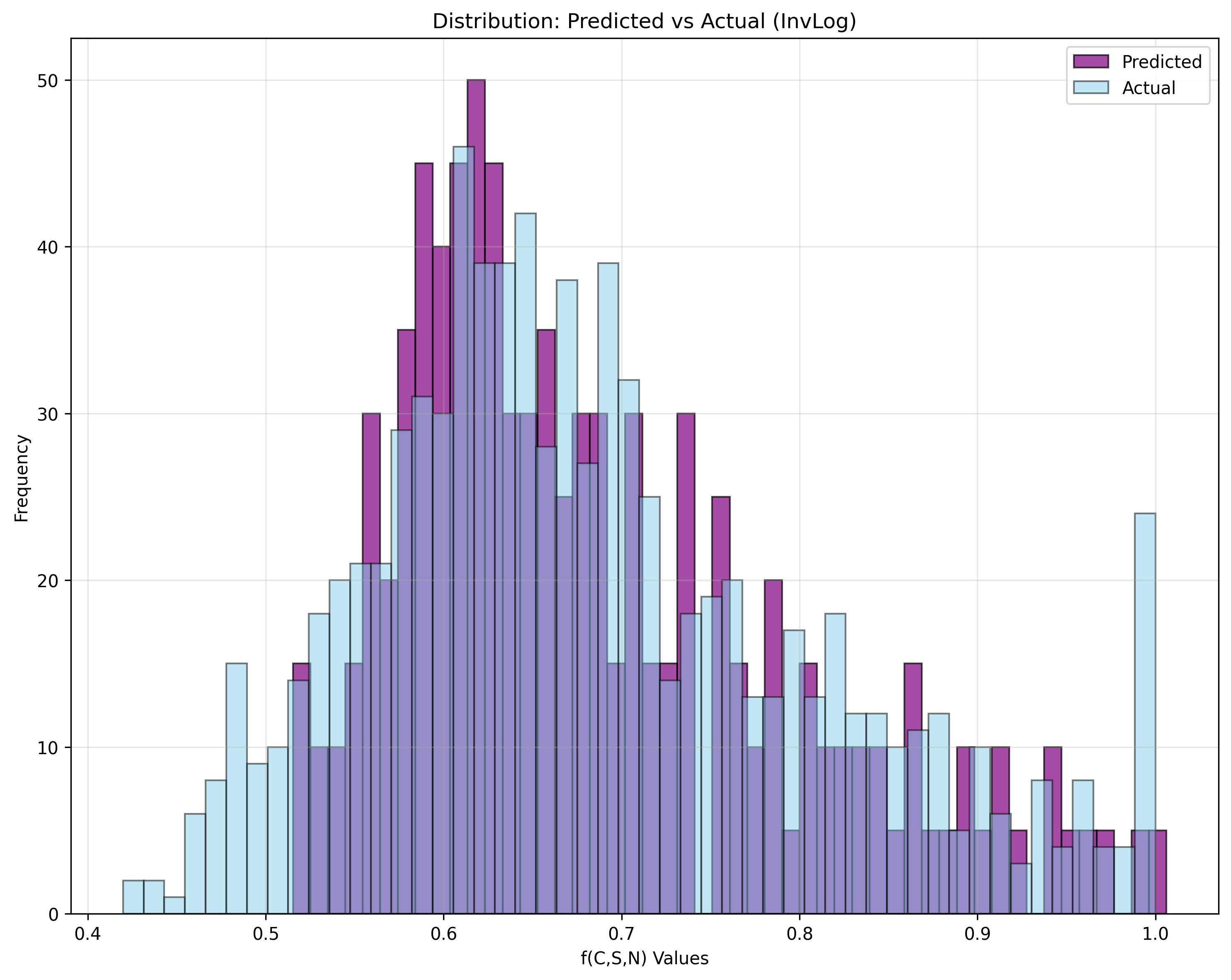}
\caption{Distribution comparison between predicted and actual accuracy values. Similar central tendencies and dispersion confirm the model captures the overall data structure.}
\label{fig:pred_actual_dist}
\end{figure}

\begin{figure}[H]
\centering
\includegraphics[width=0.8\textwidth]{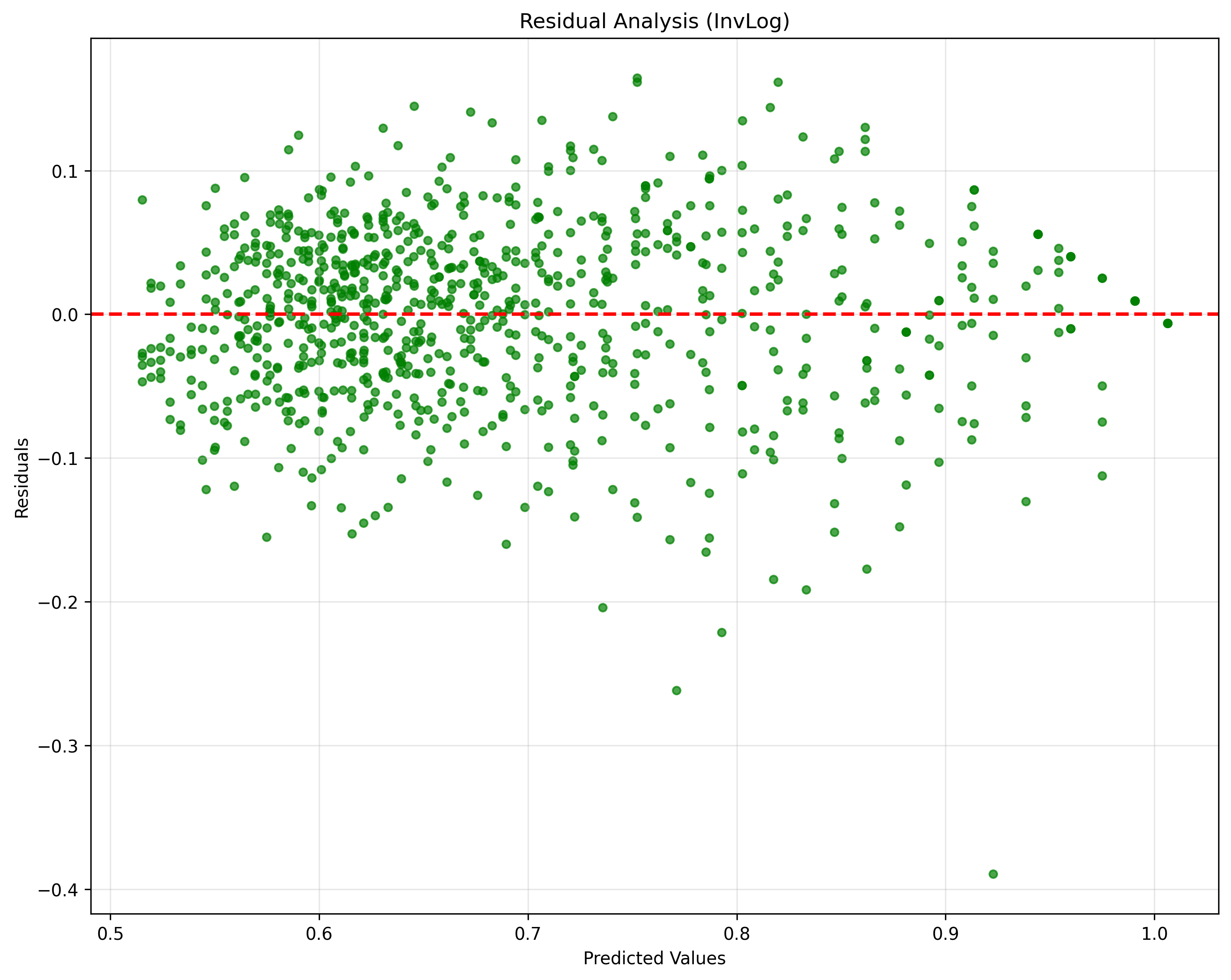}
\caption{Residual analysis showing distribution of prediction errors across accuracy values. Relatively uniform spread indicates good model fit without systematic biases.}
\label{fig:residual_analysis}
\end{figure}

\begin{figure}[H]
\centering
\includegraphics[width=0.8\textwidth]{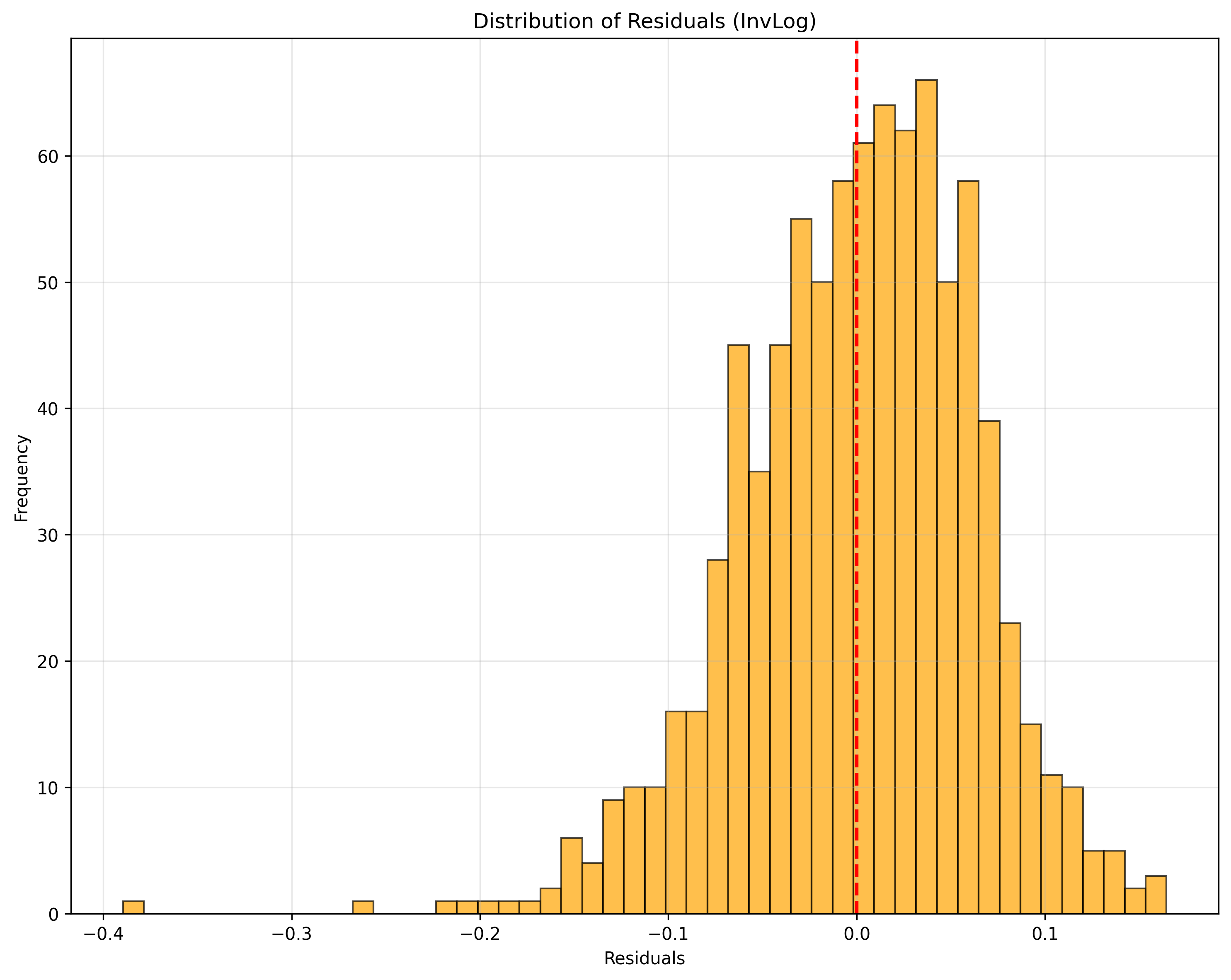}
\caption{Distribution of model residuals. Approximately normal distribution centered near zero confirms random, unbiased prediction errors.}
\label{fig:residual_dist}
\end{figure}

\end{appendices}

\end{document}